\documentclass{epl}

\title{Instability driven formation of domains in the intermediate state of type-I superconductors}
\author{V. Jeudy \and C. Gourdon}
\institute{Institut des NanoSciences de Paris - Universit$\acute{e}$s~Paris~6~et~7, CNRS~UMR~75-88\\
Campus Boucicaut, 140 rue de Lourmel, 75015 Paris - France }


\begin{document}

\maketitle

\begin{abstract}
The formation of normal-state domains in type-I superconducting indium films
is investigated using the high resolution magneto-optical imaging technique. The observed patterns consist of coexisting circular and lamellar normal-phase domains surrounded by the superconducting phase. The distribution of domain surface areas is found to exhibit a threshold, above which only the lamellar shape is observed. We show that this threshold coincides with the predicted critical surface area for the elongation instability of the circular shape. The partition of the normal phase into circular and lamellar domains is determined by the combined effects of the elongation instability and the penetration of magnetic flux by bursts at the early stage of pattern formation. It is not governed by mutual interactions between domains, as usually assumed for self-organized systems.
\end{abstract}

\section{Introduction}

A spontaneous phase separation into domains is encountered in a large number of systems including magnetic fluids~\cite{elias,cebersmaiorov,cebers}, Langmuir monolayers confined at air-water interface~\cite{mcconnell}, ferro- and ferrimagnetic layers~\cite{seul,hubertshafer}, adsorbates on a metal substrate \cite{plass}, and type-I superconductors in the intermediate state (IS)~\cite{huebener1}. The formation of domains originates from the balance between the short-range attractive interaction associated with the interfacial tension between the two phases and long-range interactions. The regular shapes of the domains are generally not stable due to the effect of long-range interactions: lamellar domains present undulation or peristaltic instabilities~\cite{cebers,seul,faber,dorsey,reisin}, circular domains (bubbles) may elongate and produce finger structures~\cite{thiele,seul-sammon,cebersmaiorov,mcconnell}. A crucial point for understanding the dynamics of domain patterns is to determine the contribution of these instabilities to the morphogenesis of domains~\cite{langer,goldstein,cebers,jackson,jagla,muratov}. 

In type-I superconductors, domain patterns are observed in film samples submitted to a perpendicular magnetic field. The IS pattern consists of coexisting normal-state (NS), flux-bearing domains, and superconducting (SC) domains. By analogy with magnetic fluids, the instability of bubble domains was proposed as the mechanism of formation of ramified structures~\cite{goldstein}. However, the progressive transformation of a circular domain into a ramified structure, which is encountered in ferromagnetic films~\cite{thiele} and in magnetic fluids~\cite{cebersmaiorov}, has not been reported in type-I superconductors up to now. This raises the question of how unstable NS bubbles can be produced in SC systems. The stability limit for the circular bubble depends on its diameter and on the magnetic Bond number $B_m$ that characterizes the ratio of the magnetic energy and the interface energy~\cite{langer}. Tuning the external field changes $B_m$ in magnetic fluids~\cite{langer} and the diameter of bubbles in ferromagnetic systems~\cite{hubertshafer}. This opens the possibility to drive the system sufficiently out of equilibrium to bring bubbles above their stability limit. In contrast, for superconductors, increasing the external field leaves unchanged both $B_m$ and the diameter of the NS bubbles in the IS pattern~\cite{jeudy,cebers_gourdon_jeudy}. 

Bubble instability could occur during the penetration of the magnetic flux into the superconductor. In type-I superconductors, the early penetration of magnetic flux proceeds by magnetic flux bursts from the sample edges~\cite{bokil,chimenti,jeudyjung,castro}. Studies of bursts remain beyond the scope of experimental investigation due to their high propagation velocity ($\approx$ 1m/s) \cite{chimenti}. Therefore, little is known about their characteristics. Whether the domains produced by the bursts are sufficiently out of equilibrium to overstep the bubble stability limit remains a fully open question.

This letter presents an analysis of the volume and shape of NS domains. A clear correlation between the surface area and the shape of domains is revealed. The threshold area, above which no more cylindrical domains are observed, is shown to correspond to the elongation instability limit. We discuss the connection between this instability and the penetration by flux bursts. Finally, the contributions of these processes to the formation of domains in type-I superconductors are emphasized. 

\section{Observation of domains in type-I superconductors}
The IS flux pattern was studied by high-resolution Faraday microscopy imaging, a technique that probes the normal component of the local induction at
the top surface of a superconductor. Experimental details are given elsewhere~\cite{cebers_gourdon_jeudy}. Indium films with thickness $d=$2.2 and 10.0
$\pm$ 0.1 $\mu$m were grown by evaporation directly onto a magneto-optic layer~\cite{gourdon}. A $112\pm 1$ $\mu m$ thick In sample was cut out from a GoodFellow 99.9 \% purity annealed foil and placed against a magneto-optic layer. The critical temperature and critical field for indium are $T_c$=3.41 K and $H_c$($T$=0)=28.2 mT, respectively. $H_{c}(T)$ was assumed to follow a Bardeen-Cooper-Schrieffer temperature variation: $H_{c}(T)=H_{c}(0)(1-(T^2/T_c^2))$. The samples were immersed into superfluid helium at temperature $T\leq$ 2 K. They were first zero-field cooled then subjected to field cycles. The external field $H_0$ was applied perpendicularly to the sample.

\begin{figure}[htpb]
\onefigure[width=0.70\textwidth]{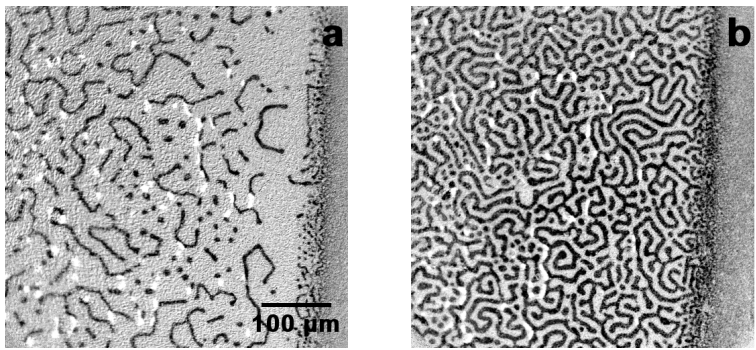}
\caption{Typical IS patterns in an indium sample with thickness $d$=10 $\mu$m for increasing values of the applied magnetic field. Images (a) and (b) correspond to $h$=0.105 and $h$=0.345, respectively ($T=$1.85 K). Flux-bearing NS domains appear in black and have circular or lamellar shapes. The edge of the sample is along the right edge of the image. The few white domains correspond to magnetic flux which was trapped at zero field~\cite{gourdon}.}
\label{ISpattern}
\end{figure}

Figure~\ref{ISpattern} shows two typical IS patterns obtained close to the edge of the sample when the field is increased.  When the reduced applied field $h=H_0/H_c(T)$ is sufficiently large, the density of domains is homogeneous even close to the edge (fig.~\ref{ISpattern}b, $h$=0.345). This is not the case at low field value (fig.~\ref {ISpattern}a, $h$=0.105) where the domain pattern is separated from the sample edge by a full diamagnetic band about 50 $\mu$m wide. This diamagnetic band is the signature of the geometrical barrier that controls the penetration of magnetic flux. This penetration from the edge to the inner part of the sample occurs sequentially, by bursts, each carrying a small amount of flux~\cite{jeudyjung,castro,chimenti,bokil}.

The two typical shapes of NS domain are clearly seen in fig.~\ref{ISpattern}: almost circular domains (bubbles) and lamellar domains. The shape of domains is usually assumed to be determined by the interface energy and the interactions between domains, as for example in ferromagnetic films~\cite{hubertshafer}. However, in superconductors, the penetration of  magnetic flux by bursts may play a predominant role in the first stage of formation of IS patterns. In order to get more insight into this point, we performed a statistical analysis of the NS domain shape.  
%
%
%
%
\section{Analysis of domain shapes}
\begin{figure}[htpb]
\onefigure[width=0.48 \textwidth]{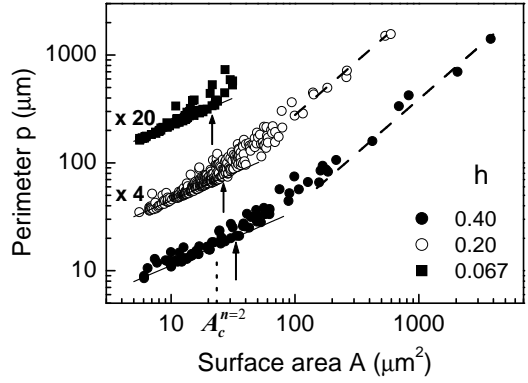}
\caption{Typical correlation between the perimeter $p$ of the domains and their surface area $A$, shown here for a 2.2 $\mu$m thick indium sample ($T=$1.815 K). The data, reported for three values of the increasing applied magnetic field $h=H_0/H_c$ are displayed in a log-log plot. For $A<$100 $\mu$m$^2$, the solid lines correspond to $p=2\sqrt{\pi A}$, \textit{i.e.} to circular domains. For $A>$100 $\mu$m$^2$, the dashed lines correspond to $p=2A/w$, \textit{i.e.} to lamellar domains much longer than wide, where $w$ is the calculated lamella width ($B_m=$0.73). The arrows indicate the threshold area value for which the domain shape starts to deviate from the circular one. The critical area for the elongation instability $A_c^{n=2}$ is indicated by the dotted line.}
\label{PerimAir}
\end{figure}

As can be qualitatively observed in fig.~\ref{ISpattern}, circular domains seem to have a smaller surface area than lamellar ones. 
In order to discriminate between bubbles and lamellae in a quantitative manner, the domain perimeter $p$ was systematically measured as a function of their surface area $A$. Typical results obtained for different values of $h$ are presented in fig.~\ref{PerimAir} for the 2.2 $\mu$m thick sample. The maximum value of $A$ increases noticeably with $h$. The weak dispersion of data indicates that $p$ and $A$ are strongly correlated.  Moreover, the data points, shown in a log-log plot, are essentially grouped along two straight lines. 
For the lowest values of $A$ (A $<$ 100 $\mu$m$^2$), the solid lines in fig. \ref{PerimAir} correspond to the equation of a circle $p=2\sqrt{\pi A}$. The agreement with the experimental data simply indicates that domains with small area have a circular shape. The dashed lines plotted in fig.~\ref{PerimAir} for large areas (A $>$ 100 $\mu$m$^2$) correspond to the equation $p=2A/w$, which is valid for long lamellae. For these plots, $w$ is the lamella equilibrium width calculated in the framework of the
``constrained current-loop'' model~\cite{cebers_gourdon_jeudy}. The good agreement with the experimental data shows that large area domains have a lamellar shape. We conclude that the shape of  a domain, either circular or lamellar, is determined by its surface area. 
%
%
%
%
%

\begin{figure}[htpb]
\onefigure[width=0.75 \textwidth]{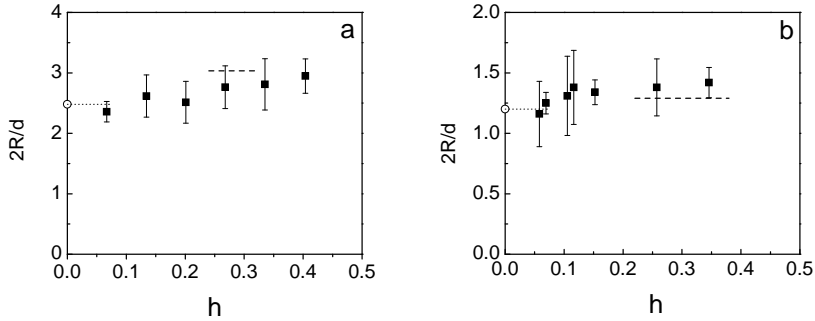}
\caption{Threshold diameter $2R_{th}/d$ (black squares) as a function of the applied magnetic field $h=H_0/H_c$ obtained for the 2.2 $\mu$m (a) and the 10 $\mu$m (b) thick indium samples. The empty circle, with the dotted line as a guide for the eye, corresponds to the critical diameter $2R_c/d$ for isolated circular domains predicted by eq.~\ref{stabilitilimit} ($B_m=d/2\pi\Delta(T)$=0.73 for $d$=2.2 $\mu$m and 3.2 for $d$=10 $\mu$m, respectively. $\Delta (T)$ is assumed to follow the empirical law $\Delta (T) = \Delta (0) /\sqrt{1-(T/T_c)}$ with $\Delta (0)$= 0.33 $\mu$m~\cite{sharvin}). The dashed lines correspond to the equilibrium diameter of circular domains calculated at the transition from the bubble hexagonal phase to the lamellar phase. The line length represent the field range in which the energies of the two structures differ by less than 0.3 \%.}
\label{AirH}
\end{figure}

Moreover, as indicated by arrows in fig.~\ref{PerimAir}, there is a threshold value for A, above which the circular shape is absent from the IS pattern. We define precisely this threshold $A_{th}$ as the value of $A$ above which no more data point is found on the line $p=\sqrt{4\pi A}$. We define here the reduced threshold diameter of bubbles $2R_{th}/d$, where $R_{th}=\sqrt{A_{th}/\pi}$. The variation of $2R_{th}/d$ with $h$ is shown in fig.~\ref{AirH} for sample thicknesses $d=$2.2 and 10.0 $\mu$m. When $h$ is raised from 0 to 0.4, the volume fraction of the sample occupied by NS domains increases from $0$ to $0.5$. However $2R_{th}/d$ is found to increase by only $20 \% $ and $10 \%$ for the $2.2$ and the $10.0$ $\mu$m samples, respectively. This means that $2R_{th}/d$ only weakly depends on mutual interactions between domains, and suggests a comparison of the threshold diameter with the stability limit of an isolated circular domain~\cite{cebersmaiorov,langer}.
%
%
%
%
\section{Stability limit for the circular shape}
In order to determine the critical diameter above which the circular shape becomes unstable, we consider the linear dynamics near the circular initial condition. The calculation is performed in the framework of the ``constrained current-loop'' model~\cite{cebers_gourdon_jeudy}. Let us assume that the applied magnetic field $H_0$ produces a set of $N$ NS circular domains of radius $r$, each bearing an induction $h_n$. The constraint of flux conservation is written $H_0=\rho_n h_n$ where $\rho_n=N \pi r^2 d/V$ is the volume fraction of the normal phase in a sample of volume $V$. In the limit of non-interacting bubbles ($H_0\rightarrow$0), the free energy per unit volume of the system is 
%
%
\begin{equation}
\label{energy}
   F=F_s+ \frac{1}{8 \pi}\left(\rho_n H_c^2+\frac{H_0^2}{\rho_n}\right)+
   \frac{\rho_n}{d}\left(\frac{2 \sigma_{NS} d}{r}+\frac{h_n^2 r}{3 \pi^2}\right)\;,      
\end{equation}
%
%
where $F_s$ is the free energy density of the SC state. The two terms in the first parenthesis are the ``bulk'' terms corresponding to the condensation energy and the magnetic energy respectively. The first term in the second parenthesis is the interface energy where the surface tension $\sigma_{NS}$ is equal to $\Delta H_c^2/8 \pi$, with $\Delta$ the interface energy parameter. The last term is the correction of the magnetic energy due to the curvature of the flux lines in the free space on both sides of the sample. It can be calculated using the Biot-Savart interaction of the screening current encircling NS domains~\cite{goldstein, cebers_gourdon_jeudy}. Minimizing $F$ with respect to $\rho_n$ and $r$ yields $h_n=H_c$ and the equilibrium bubble diameter. 

\begin{figure}[htpb]
\onefigure[width=0.48 \textwidth]{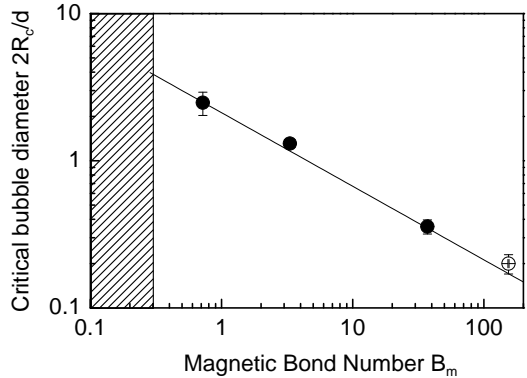}
\caption{Reduced critical bubble diameter $2R_c/d$ as a function of the magnetic Bond number $Bm$ shown in a log-log plot. The filled circles represent the threshold diameter in the limit $h\rightarrow$0 for In samples. The empty circle is
reported from ref. \cite{farrell} for Hg. The shaded region corresponds to type-II superconductivity for very thin samples. The solid line represents the critical diameter of an isolated bubble predicted by eq. \ref{stabilitilimit}. A good agreement is obtained over more than two orders of magnitude of $B_m$ without any adjustable parameter.}
\label{RvsBm}
\end{figure}

For the calculation of the stability limit, the domain wall motion is supposed to result from the balance between the generalized force $\hat{\bf{e}}_r \delta F/ \delta \bf{r}$ and a local viscous force $-\eta \hat{\bf{e}}_r \partial \bf{r}/\partial \textit{t} $~\cite{langer,goldstein}. The resulting equation of motion is 
%
\begin{equation}
\label{equamotion}
   -\eta \hat{\bf{e}}_r\frac{\partial \bf{r}}{\partial t} = \kappa \sigma_{NS} +\frac{H_c^2}{6 \pi^2 d} \bf{r}\:\hat{\bf{e}}_r\;,  
\end{equation}
where $\kappa$ is the interface curvature. Consider small deviations $\xi$ in the radius of the circle, parametrized by the polar angle $\varphi$: $\bf{r}(\varphi,\textit{t})=\left[R_0+ \xi(\varphi,t)\right]\hat{\bf{e}}_r(\varphi)$ and suppose that the perturbation is in the form $\xi(\varphi,t)\approx exp(\sigma_n t)cos(n\varphi)$. The growth rate of mode $n$ is then given by 
\begin{equation}
\label{growth-rate}
\sigma_n = \left[(1-n^2)/R_0^2+ 8 B_m/3 d^2\right]\sigma_{NS}/\eta\;,
\end{equation}
where the magnetic Bond number $B_m$ is equal to $d/2 \pi \Delta(T)$. Note that the same result is obtained for magnetic fluids in the limit $d\rightarrow\infty$ \cite{cebers,langer}. The stability limit is reached for $\sigma_n=0$: 
%
\begin{equation}
\label{stabilitilimit}
  \frac{2 R_c^n}{d} = \sqrt{\frac{3(n^2-1)}{2B_m}}    
\end{equation}
%
The reduced critical diameter $2R_c^{n=2}/d$ predicted for the elongation instability is plotted in fig.~\ref{AirH} and compared to the threshold diameter $2R_{th}/d$, measured for the $2.2$ and the $10.0$ $\mu$m thick samples. Within the experimental error bars, $2R_c^{n=2}/d$ coincides with the extrapolation of $2R_{th}/d$ for $h \rightarrow$ 0. The variation of the stability limit with increasing inter-domain interactions remains beyond the scope of this paper~\cite{jackson2003}. However, one can compare the maximum threshold, $2R_{th}/d$ at $h\approx$ 0.4, to the predicted equilibrium bubble diameter for which the free energies of an hexagonal array of bubbles and a 1D-array of lamellae are equal (fig.~\ref{AirH}). The good quantitative agreement indicates that the measured variations of the threshold $2R_{th}/d$ with $h$ are consistent with the two limits for the critical diameter: for $h$ close to zero or close to the bubble-lamellae transition field. Moreover, $2R_{th}/d (h \rightarrow 0)$ was systematically measured for different sample thicknesses. The results are shown in fig.~\ref{RvsBm}. The same figure also displays the result
obtained for Hg in ref.~\cite{farrell} for $\rho_n\approx 0.3-0.4$ with $\Delta (0)$ = 0.084 $\mu$m~\cite{jeudy}. All the data gather on a single master curve indicating that $2R_c/d$ and $B_m$ are appropriate reduced variables. A very good quantitative agreement is obtained with the prediction of eq. \ref{stabilitilimit}, without any adjustable parameter. Note that the magnetic Bond number $B_m$ is varied by more than two orders of magnitude (0.72-150). 
This demonstrates that the shape (circular or lamellar) of the NS domains is controlled by the elongation instability limit of the circular shape.
%
%
%
\section{Discussion}
This result raises the general question of the role of the elongation instability in the formation of patterns in type-I superconductors. It was proposed in Ref. \cite{goldstein} that the formation of lamellar structures could originate from the instability of bubbles. A careful examination of NS domains in the $2.2$ and the $10.0$ $\mu$m thick samples indicates that the NS bubbles start to disappear only when $h$ reaches $\approx 0.3-0.4$. Above this limit, the disappearance of bubbles is found to result from their fusion with other bubbles or with already existing lamellae. Below $h \approx 0.3-0.4$, the increase of $h$ does not change NS bubble surface area. This is a consequence of the conservation of magnetic flux ($\Phi \approx H_c A$) in a NS domain isolated in the SC matrix~\cite{jeudy}. Therefore $A$ cannot increase beyond the stability limit.  We conclude that, unlike magnetic fluids~\cite{cebersmaiorov} and ferromagnetic films~\cite{thiele}, the early formation of lamellar structures in type-I superconductors does not result from the instability of bubbles already present in the IS pattern. 

We now address the question of the contribution of flux bursts to the shape of NS domains. Isolated NS domains appearing at low field are produced by magnetic flux bursts. The mechanisms of production, as well as the distribution of amplitude of the bursts, remain to a large extent unexplored. 
When $h$ is increased from zero, NS domains first appear as bubbles. With further increase of the field, lamellar domains appear~\cite{cebers_gourdon_jeudy}. As seen in fig.~\ref {PerimAir}, this leads, for the 2.2 $\mu$m sample, to a large increase of the fraction of lamellae between $h=0.067$ and $h=0.2$. Those lamellae do not arise from the fusion of bubbles since the latter start to disappear only when $h$ reaches $\approx 0.3-0.4$: they are directly produced by flux bursts. The penetration of magnetic flux bursts thus results in the production of both bubbles and lamellae.

Furthermore, images of the IS pattern show a leveling-off of the number of NS bubbles produced by flux bursts when the external field is increased~\cite{cebers_gourdon_jeudy}. For the $2.2$ and the $10.0$ $\mu m$ thick films, this leveling-off occurs at $h=0.10-0.15$ while the free energy of the bubble lattice remains lower than the free energy of the lamella lattice up to $h\approx0.3$~\cite{cebers_gourdon_jeudy}. As a consequence, the leveling-off of the number of NS bubbles is not governed by the competition between the interface energy and long-range magnetic interactions, as assumed in equilibrium mean-field models~\cite{muratov,cebers_gourdon_jeudy}.

In view of the above experimental results, we propose the following scenario for the formation of NS domains. The shape of NS domains is determined during their penetration through the diamagnetic band. The amplitude of flux bursts, \textit{i.e.} the surface area $A= \Phi / H_c$ of the NS domains released on the edge of the sample, increases with $h$. For small values of $h$, the released domains have a surface area smaller than the critical area for the elongation instability $A_c^{n=2}$: they become bubbles. With further increase of $h$, the surface area of the largest released domains exceeds $A_c^{n=2}$: they become lamellae. The appearance of new bubbles stops for a field value $h$ such that the surface area of all the released domains is larger than $A_c^{n=2}$.
Therefore the NS domain shape depends on the distribution of amplitudes of flux bursts and its evolution with the external field with respect to the critical area for the elongation instability. We emphasize that the process proposed here for domain formation concerns only the low field range. With increasing field, NS domains become connected to the sample edge and the growth of the normal phase proceeds by continuous penetration of the magnetic flux~\cite{cebers_gourdon_jeudy}.

\section{Conclusion}
In conclusion, our study shows clear evidence of the existence of the elongation instability of the circular shape in type-I superconductors. The role of this instability in the formation of domain pattern is clarified. At the early stage of flux penetration, the partition between bubbles and lamellae is not driven by a thermodynamical equilibrium resulting from the competition between long-range and short-range interactions. It is determined by both the bubble elongation instability limit and the distribution of amplitudes of flux bursts penetrating on the edges of the samples. 
These results explain well why almost perfect hexagonal arrays of bubbles are hardly observed in type-I superconductors while they are commonly reported in ferromagnetic films and magnetic fluids. 
We hope that this work will stimulate further theoretical work in order to get more insight into the dynamics of penetration of the magnetic flux.

\acknowledgments
The authors gratefully acknowledge stimulating and fruitful discussions with A. Cebers.

\end{document}